\documentclass[rnote]{aa} 

\usepackage{graphicx}
\usepackage{grffile}
\usepackage{txfonts}

\begin{document} 
\newcommand{\hi}{H\,I}
\newcommand{\HI}{H\,I\,\,}
\newcommand{\NHI}{$\rm{N_{\mathrm{H\,I\,}}}$}
\newcommand{\cmm}{cm$^{-2}\:$}
\newcommand{\cmmm}{cm$^{-3}\:$}
\newcommand{\kms}{km~s$^{-1}\:$}
\newcommand{\Msun}{$M_{\odot}$}
\newcommand{\gsubselector}[1]{$H_{2}$}

\title{LIMITS ON THE H I CONTENT OF THE DWARF GALAXY HYDRA II}

\author{Andrew Janzen\inst{1},
Eve M. Klopf\inst{2},
Felix J. Lockman\inst{3},
Rodolfo Montez Jr\inst{4},
Kurt Plarre\inst{5}, 
Nau Raj Pokhrel\inst{6},
Robert J. Selina\inst{7},
A. Togi\inst{8},
Mehrnoush Zomederis\inst{9}}

\institute{California Institute of Technology, 1200 E. California Blvd, Pasadena, CA 91125, USA
\and
Oregon Institute of Technology, 3201 Campus Drive, Klamath Falls, OR 97601, USA
\and
National Radio Astronomy Observatory, Green Bank, WV 24494, USA\\
\email{jlockman@nrao.edu}
\and
Physics and Astronomy Department, Vanderbilt University, PMB 401807, 2401 Vanderbilt Place, Nashville, TN 37240, USA
\and
Associated Universities Inc, Av. Nueva Costanera 4091, Suite 502, Vitacura Santiago, Chile
\and
Department of Physics, Florida International University, Miami, FL 33199, USA
\and
National Radio Astronomy Observatory, PO Box O, Socorro, NM 87801, USA
\and
Department of Physics and Astronomy, The University of Toledo, 2825 West Bancroft Street, Toledo, OH 43606, USA
\and
Department of Physics and Astronomy, York University, 4700 Keele Street, Toronto, Ontario, M3J 1P3, Canada
}

\date{Received September 11, 2015; accepted }

\abstract 
{Sensitive 21cm \HI observations have been made with the Green Bank Telescope toward the newly-discovered Local Group dwarf galaxy Hydra~II, which may lie within the leading arm of the Magellanic Stream.  No neutral hydrogen was detected. Our $5\sigma$ limit of M$_{\rm HI} \leq 210 $ M$_{\odot}$ for a 15 \kms  linewidth gives a gas to luminosity ratio M$_{\rm HI}$/L$_V \leq 2.6 \times 10^{-2}$ M$_{\odot}$ L$_{\odot}^{-1}$.  The limits on \HI mass and M$_{\rm HI}$/L$_V$ are typical of dwarf galaxies found within a few hundred kpc of the Milky Way.  Whatever the origin of Hydra~II, its neutral gas properties are not unusual.}


\authorrunning{Janzen et al.}
\titlerunning{H I in Hydra II}
\maketitle

\section{Introduction}
Hydra~II is a dwarf galaxy in the Local Group, newly-discovered during the Survey of the Magellanic Stellar History (SMASH) using the Dark Energy Survey Camera DECam \citep{Nidever15,Martin15}. It lies at a distance of 134 kpc from the Sun with V$_{hel} = 303$ \kms, and has  physical properties, such as a half light radius of 66 pc, a high dynamical mass-to-luminosity ratio, and a metallicity $\langle \vert Fe/H\vert \rangle = -2.02\pm0.08$, indicating that it is a dwarf galaxy and not a cluster in the Milky Way's halo \citep{Martin15, Kirby15}. It is located in the same region of the sky as the leading arm of the Magellanic Stream and has a similar velocity \citep{Nidever08, Nidever10}.  The distance to the leading arm is not known and estimates range from 21 kpc to $>100$ kpc \citep{McClure-Griffiths08, Besla12, Martin15}.  If Hydra II originated in the Magellanic Clouds it may have a considerably different history and properties than other Milky Way dwarfs \citep{Martin15}. For this reason, we made very sensitive observations of the 21cm \HI emission in Hydra~II using the Green Bank Telescope, which has been used to produce some of the most strict limits on the \HI content of other Local Group dwarf galaxies \citep{Spekkens14}.

\section{Observations and analysis}
Measurements of 21cm emission from Hydra~II were made with the 100-meter diameter Robert C.~Byrd Green Bank Telescope (GBT) of the National Radio Astronomy Observatory \footnote{The National Radio Astronomy Observatory is a facility of the National Science Foundation operated under cooperative agreement by Associated Universities, Inc.} on 2015, July 7--8 and July 17, centered at the position J2000 = $\rm{12^{h}21^{m}42.1^{s}}$ $-31^{\circ}59^{\arcmin}07^{\arcsec}$ given by \citet{Martin15}. The GBT 1.4 GHz receiver had a total system temperature toward Hydra~II of about 21 K. The GBT VEGAS spectrometer was used to measure spectra with both in-band frequency switching for 23 minutes and position-switching to a reference location displaced $\pm 5^m 30^s$  in right ascension from the source for 110 minutes.  The $9\farcm1$ FWHM beam of the GBT at 21cm completely encompasses Hydra~II, which is estimated to have an angular  size of about $1\farcm7$ on the sky \citep{Kirby15}.

The 21cm spectra covered $\pm$2000 \kms centered on zero velocity in the local standard of rest (LSR). In the position-switched observations, emission at Milky Way velocities between -70 and +180 \kms is partially cancelled, but as  the stars in Hydra~II have a mean V$_{\rm LSR} = +301$ \kms, this does not affect our results. The data were reduced using the standard GBTIDL routines and calibrated correcting for atmospheric attenuation to produce T$_{a}^{\star}$ as a function of V$_{\rm LSR}$. Frequency-switched data were calibrated and corrected for stray radiation using the method described in \citet{Boothroyd11}. The final spectra, smoothed to a velocity resolution of 0.30 \kms from the intrinsic resolution of 0.15 \kms, are shown in Figure 1. We combine the noise limits from position- and frequency-switched spectra for a final estimated rms noise in T$_{a}^{\star}$ of 9.1 mK (4.55 mJy) in an 0.30 \kms channel. 

\begin{figure}
\centering
\includegraphics*[width=0.5\textwidth]{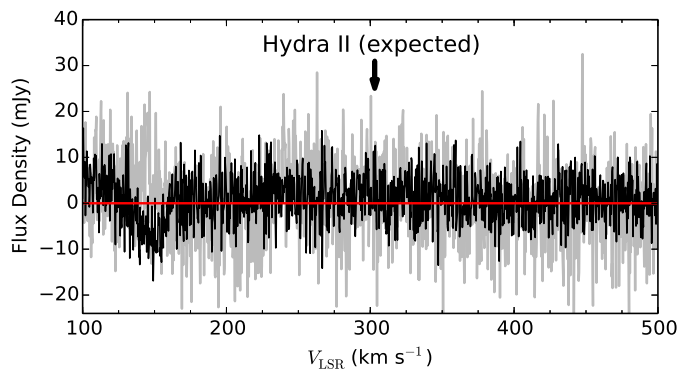}
\caption{GBT 21cm \HI spectra toward Hydra II over the velocity range appropriate to that galaxy. The grey and black are the frequency- and position-switched spectra, respectively. Their differing noise level is a consequence  of the difference in exposure times: 23 minutes for frequency switching and 110 minutes for position switching. The combined rms noise is 9.1 mK in an 0.30 \kms channel.  The dip in the position-switched spectrum near 150 \kms results from partial cancellation of Milky Way emission.}
\end{figure}

\begin{figure}
\centering
\includegraphics*[width=0.5\textwidth]{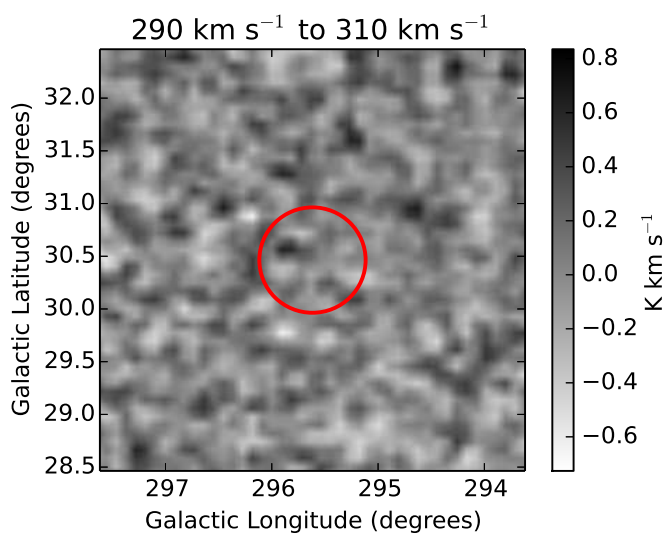}
\caption{The integrated 21cm \HI emission in the field of Hydra~II from the GASS survey \citep{McClure-Griffiths09, Kalberla10} over 20 \kms around the stellar velocity of Hydra~II, whose location at $\ell,b = 295\fdg62 +30\fdg46$ is marked with the circle. No significant emission is detected. The brightest \HI within a degree of the galaxy is consistent with  a $2.3\sigma$ noise feature.}
\label{fig:hydragass}
\end{figure}

No signal from \HI associated with Hydra~II is detected.  The $5\sigma$ limit on the velocity-integrated flux density over 15 \kms  is $ 5 \times 4.55 \times 0.3 \times \sqrt{15/0.30} =  48.3 $ mJy \kms. The \HI mass within the beam is given by 
\begin{equation}
M_{\rm HI} \  {(M_{\odot})} = 2.36 \times 10^{5} \ D^2 \int S dv\ 
\label{eqn:atmass}
\end{equation}
\citep{Roberts75}, 
where D is the source distance in Mpc, and $\int S \mathrm{d}v$ is the velocity integrated flux in Jy \kms. We assume an \HI linewidth of 15 km s$^{-1}$, consistent with that used in the analysis of other local dwarfs \citep{Spekkens14}. At the 0.134 Mpc distance of Hydra~II, our measurements give a   5$\sigma$ limit  M$_{\rm H\,I} \leq  210$ M$_{\odot}$.
 
To explore whether there might be some \HI associated with Hydra~II but displaced somewhat in position or velocity from the stars, we integrated the GASS \HI survey, made with the Parkes radio telescope at an angular resolution of 14\farcm4   \citep{McClure-Griffiths09,Kalberla10} over 20 \kms around the velocity of the galaxy. Figure \ref{fig:hydragass} shows the resulting noise map. No \HI is detected. The brightest GASS pixel within one degree of Hydra~II has an integrated intensity of 0.7 K-\kms, or $2.3\sigma$ of the measured noise.  Over the area of the Parkes telescope beam anywhere in this field, the 5$\sigma$ limit on  M$_{\rm H\,I}$  is 9350 M$_{\odot}$.

\section{Discussion}
\begin{figure}
\centering
\includegraphics*[width=0.5\textwidth]{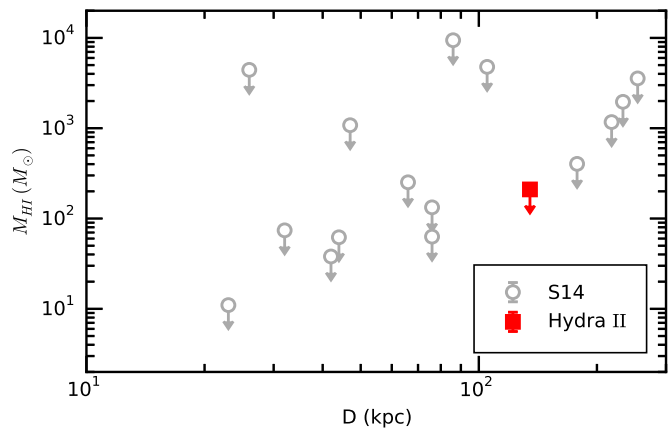}
\caption{5$\sigma$ \HI mass limits of local dwarf galaxies from \citet{Spekkens14} (S14) are shown in open symbols as a function of heliocentric distance, with the Hydra~II measurement as the filled symbol. The limit on M$_{\rm HI}$  of Hydra~II is consistent with that of other dwarfs near the Milky Way.}
\label{fig:hydramass}
\end{figure}

\begin{figure}
\centering
\includegraphics*[width=0.5\textwidth]{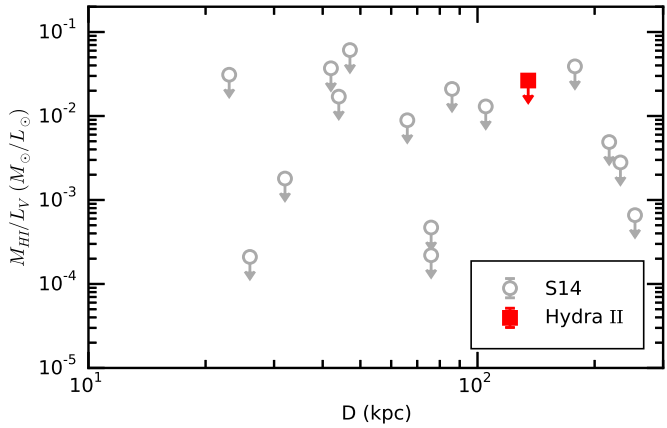}
\caption{5$\sigma$ \HI mass to L$_V$ ratios of local dwarf galaxies  from \citet{Spekkens14} (S14) are shown as a function of heliocentric distance with open symbols, and the Hydra~II measurement as a filled symbol using L$_V$ from \citet{Martin15}. The limit for Hydra~II is consistent with that of other dwarfs near the Milky Way.}
\label{fig:hydra_M_L}
\end{figure}

We do not detect any \HI directly toward Hydra~II, nor is there any evidence of significant \HI nearby that galaxy, in position or velocity. Our  $5\sigma$ limit  M$_{\rm HI} \leq  210$ M$_{\odot}$ shown in Figure ~\ref{fig:hydramass}, is consistent with values derived for other Local Group dwarf galaxies that lie within the $\approx 300$ kpc virial radius of the Milky Way \citep{Blitz00,Grcevich09,Spekkens14,Westmeier15}. Likewise, the derived \HI mass--luminosity limit (using L$_V$ from \citet{Martin15}) of M$_{\rm HI}$/L$_V \leq 2.6 \times 10^{-2}$ M$_{\odot}$ L$_{\odot}^{-1}$ is also similar to that found for dwarfs near the Milky Way, as shown in Figure ~\ref{fig:hydra_M_L}. Dwarf galaxies at larger distances from the Milky Way typically have M$_{\rm HI}$/L$_V \approx 1$ \citep{Connachie12}. 
 
The lack of  detectable \HI in nearby dwarf galaxies is usually ascribed to tidal or ram pressure stripping in the Milky Way's hot halo \citep{Blitz00,Mayer06,Grcevich09,Gatto13}. Presumably, similar processes have removed the gas from Hydra~II. In this regard, Hydra~II's location  near the leading arm of the Magellanic Stream has apparently not affected its neutral Hydrogen properties as compared with other dwarfs at similar distances.  

\begin{acknowledgements}
We thank K. Spekkens for useful discussions and Richard Prestage for comments on the manuscript. These measurements were performed as part of  the 2015 NRAO/NAIC Single Dish Summer School program.
\end{acknowledgements}


\bibliography{ref}
\bibliographystyle{aa}
\listofobjects {Hydra II}
\end{document}